\documentclass[conference]{IEEEtran}
\usepackage{amsmath}
\usepackage{amssymb}
\usepackage{amsfonts}
\usepackage{cite}
\usepackage{csquotes}
\usepackage{graphicx}

\usepackage{makecell}
\usepackage{multirow}
\usepackage{url}

\usepackage{xcolor}
\def\BibTeX{{\rm B\kern-.05em{\sc i\kern-.025em b}\kern-.08em T\kern-.1667em\lower.7ex\hbox{E}\kern-.125emX}}

\DeclareMathSymbol{:}{\mathord}{operators}{"3A}

\begin{document}

\title{Combined Neural Network-based Intra Prediction and Transform Selection}

\author{
Thierry~Dumas, Franck~Galpin, Philippe~Bordes\\ \IEEEauthorblockA{\textit{Interdigital}, Rennes, France\\thierry.dumas@interdigital.com, franck.galpin@interdigital.com, philippe.bordes@interdigital.com}
}

\maketitle

% Abstract.
% !TeX root = combined_neural_network.tex

\begin{abstract} \label{section:abstract}
The interactions between different tools added successively to a block-based video codec are critical to its rate-distortion efficiency. In particular, when deep neural network-based intra prediction modes are inserted into a block-based video codec, as the neural network-based prediction function cannot be easily characterized, the adaptation of the transform selection process to the new modes can hardly be performed manually. That is why this paper presents a combined neural network-based intra prediction and transform selection for a block-based video codec. When putting a single neural network-based intra prediction mode and the learned prediction of the selected LFNST pair index into VTM-8.0, $-3.71\%$, $-3.17\%$, and $-3.37\%$ of mean BD-rate reduction in all-intra is obtained.
\end{abstract}

\begin{IEEEkeywords}
Transform signaling, intra prediction, neural networks, Versatile Video Coding.
\end{IEEEkeywords}

% Introduction.
% !TeX root = combined_neural_network.tex

\section{introduction} \label{section:introduction}
In a block-based video codec featuring multiple transforms, the signaling of the selected inverse transform to be applied to a block of reconstructed transform coefficients at the decoder can take two forms. In the first form, a.k.a explicit signaling, the encoder writes to the bitstream the selected transform index. This way, the decoder identifies the selected inverse transform by reading its index from the bitstream. In the second form, a.k.a implicit signaling, the decoder derives the selected transform index from available information.

Given that the implicit transform signaling does not spend any bit, it makes particular sense in terms of rate-distortion when the used available information correlates with the efficiency of the transforms at compacting the residual block energy into few transform coefficients. This is well illustrated by the Low Frequency Non-Separable Transform (LFNST) \cite{low_frequency_non_separable} in Versatile Video Coding (VVC). During the LFNST training, the $67$ VVC intra prediction modes are first divided into groups. Then, for each group, a pair of LFNST matrices is trained on blocks of primary transform coefficients, each of them arising from the application of the DCT2-DCT2 to the residue resulting from the intra prediction of an image block via a mode of this group. Consequently, in the VVC decoder, if a block of reconstructed transform coefficients uses LFNST, its selected intra prediction mode index (the used available information) maps to the index of the pair of LFNST matrices trained on data generated via this mode\footnote{See "g$\_$lfnstLut" at \url{https://vcgit.hhi.fraunhofer.de/jvet/VVCSoftware_VTM/-/blob/master/source/Lib/CommonLib/RomLFNST.cpp}}, i.e. the one probably having the highest energy compaction efficiency in the current case. Note that the selected LFNST matrix index among the chosen pair is explicitly signaled.

Unfortunately, an implicit transform signaling deriving from the selected intra prediction mode index can hardly benefit to an intra prediction mode added to the block-based video codec afterwards as no straightforward correlation exists between the new mode index and the energy compaction efficiencies of the transforms of interest. This happened to the implicit LFNST signaling when introducing the Matrix-based Intra Prediction (MIP) \cite{affine_linear_weighted_intra} modes in VTM-5.0. To correct this, a fixed mapping from each MIP mode index to one of the $67$ VVC modes indices considered by the implicit LFNST signaling appeared in VTM-5.0. Note that, since VTM-6.0, the mapping from each MIP mode index to $0$ has replaced the fixed mapping in VTM-5.0, thus showing its low rate-distortion impact.

More critically, if the new intra prediction mode is made of deep neural networks, a solution similar to the above-mentioned mapping from the new mode index to an intra prediction mode index considered by the implicit transform signaling becomes unfeasible for two reasons. Firstly, similarities between the prediction of a block via the new non-linear mode and the predictions of this block via the existing linear modes cannot sometimes be found, preventing the association of modes indices. Secondly, as the characteristics of the deep neural network intra prediction of a block, e.g. horizontal propagation of the pixel intensities from decoded samples around this block into the predicted block, depend on the decoded neighboring samples \cite{fully_connected_network_based, progressive_spatial_recurrent_neural, fully_neural_network_mode, multi_scale_convolutional_neural}, this kind of mapping cannot be fixed at the encoder and the decoder.

Alternatively, this paper proposes to learn the selection of the transform index from an intermediate representation of the neural network prediction of the block. This way, for a block predicted via the neural network-based mode, the implicit transform signaling adapts to the decoded samples surrounding this block and the non-linear prediction function. Note that, in \cite{cnn_based_transform_index}, a neural network predicts the selected transform index from the block of quantized transform coefficients. Differently, our approach aims at integrating into a block-based video codec both a new coding tool, i.e. a neural network-based intra prediction mode, and a complement to the implicit transform signaling now depending on this new coding tool.

When inserting a single additional neural network-based intra prediction mode and the learned prediction of the selected LFNST pair index into VTM-8.0, $-3.71\%$, $-3.17\%$ and $-3.37\%$ of mean BD-rate reduction in all-intra is reported.

% Neural network-based trasform selection.
% !TeX root = combined_neural_network.tex

\section{Neural network-based transform selection} \label{section:neural_network_based_transform}
\begin{figure*}
	\centering
	\includegraphics[width=0.96\linewidth]{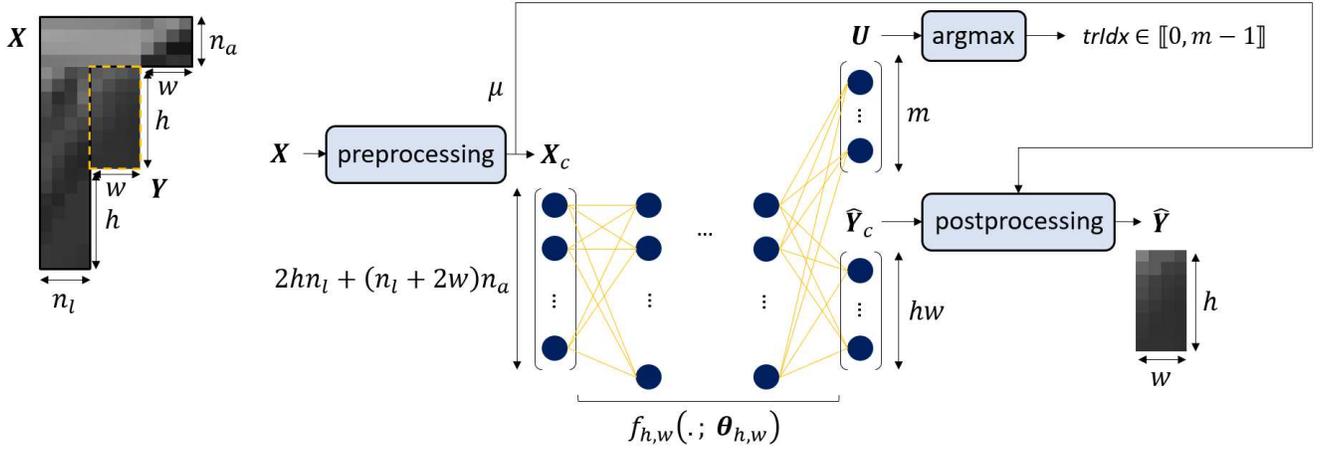}
	\caption{Prediction $\hat{\mathbf{Y}}$ of a $w \times h$ block $\mathbf{Y}$ and inference of the selected transform index \textit{trIdx} from the block context $\mathbf{X}$ via the neural network $f_{h, w} \left( \; . \; ; \boldsymbol{\theta}_{h, w} \right)$. Here, the implicit transform signaling features the selection of one transform among $m$ possible transforms. $\mu$ gathers pre-processing variables needed by the post-processing step, see an example in Section \ref{subsection:rate_distortion_analysis_of}.}
	\label{figure:intra_prediction_implicit_signaling}
	\vspace{-4.0mm}
\end{figure*}
The method developed in this paper targets the integration of neural network-based intra prediction modes and the learned transform index selection into a block-based video codec. Two factors orient our study towards considering a single neural network-based mode. Firstly, the addition of a single neural network-based intra prediction mode to VVC, this mode having a relatively small signaling cost with respect to those of the $67$ VVC intra prediction modes and the MIP modes, has already shown significant rate-distortion gains \cite{iterative_training_of_neural}. Secondly, when the index returned by a learned mapping is used for predictive coding, see Section \ref{subsection:signaling_in_vvc}, the extra encoder running time caused by the additional tests of transforms remains small only in the case of a single neural network-based mode. Despite this orientation, our approach can be easily generalized to multiple new neural network-based intra prediction modes.

In this single neural network-based intra prediction mode, blocks of size $w \times h$ in the codec are predicted by the neural network $f_{h, w} \left( \; . \; ; \boldsymbol{\theta}_{h, w} \right)$, parametrized by $\boldsymbol{\theta}_{h, w}$. Indeed, as a neural network for intra prediction takes the L-shape context around a block to provide the predicted block, see Figure \ref{figure:intra_prediction_implicit_signaling}, its architecture must include full-connections, making the number of neural network parameters dependent on the block size. Therefore, the single neural network-based mode contains $\text{card} \left( Q \right)$ neural networks, $Q$ denoting the set of possible pairs of block height and width in the codec.

Given the composition of the single additional mode, each of its neural network must own a different mapping to the selected transform index. More precisely, the neural network $f_{h, w} \left( \; . \; ; \boldsymbol{\theta}_{h, w} \right)$ takes a pre-processed version $\mathbf{X}_{c}$ of the context $\mathbf{X}$ of decoded samples around a $w \times h$ block $\mathbf{Y}$ to return both a prediction $\hat{\mathbf{Y}}_{c}$ of $\mathbf{Y}$ before post-processing and a vector $\mathbf{U}$ of unscaled log-probability of each transform to be selected, see Figure \ref{figure:intra_prediction_implicit_signaling}. Then, the post-processing turns $\hat{\mathbf{Y}}_{c}$ into the final prediction $\hat{\mathbf{Y}}$ of $\mathbf{Y}$ and the selected transform index $\textit{trIdx}$ according to the neural network corresponds to the position of the maximum in $\mathbf{U}$. Note that a fully-connected architecture is displayed in Figure \ref{figure:intra_prediction_implicit_signaling}. Yet, Figure \ref{figure:intra_prediction_implicit_signaling} can be adapted to a convolutional architecture like the one in Section \ref{subsection:rate_distortion_analysis_of}. Note also that the penultimate neural representation in $f_{h, w} \left( \; . \; ; \boldsymbol{\theta}_{h, w} \right)$ is chosen as input to the fully-connected layer returning $\mathbf{U}$ because, in the conditions of the experiments in Section \ref{subsection:rate_distortion_analysis_of}, it has been observed that this input yields the best accuracy of the classification of the selected transform.

% Learned transform selection for LFNST.
% !TeX root = combined_neural_network.tex

\section{Neural network-based LFNST selection} \label{section:learned_transform_selection_for}
This section applies the generic combined neural network-based intra prediction and transform index selection in Section \ref{section:neural_network_based_transform} to the implicit LFNST signaling in VVC. First, the indexing of the secondary transforms of LFNST is changed to suit the neural network-based transform index selection framework, see Section \ref{subsection:modified_indexing_of_the}. Then, Sections \ref{subsection:training_of_the_neural} and \ref{subsection:signaling_in_vvc} describe respectively the training of the neural network-based LFNST selection and the signaling of the proposed method in VVC.

\subsection{Modified indexing of the secondary transforms of LFNST} \label{subsection:modified_indexing_of_the}
The LFNST signaling mixes an implicit signaling and an explicit one. Regarding the implicit signaling, the index of the intra prediction mode selected to predict the current Coding Block (CB) determines the index of the pair of LFNST matrices among four pairs and whether the primary transform coefficients resulting from the application of the DCT2 horizontally and the DCT2 vertically to the residue of prediction are transposed, see Table \ref{table:indexing_implicit_lfnst}. Note that this implicit signaling stems from the LFNST training summarized in Section \ref{section:introduction}. Regarding the explicit signaling, \textit{lfnstIdx} $= 0$ means that the encoding/decoding of the current CB does not use LFNST. \textit{lfnstIdx} $\in \left\{ 1, 2 \right\}$ indicates the selected LFNST matrix index among the pair given by the implicit LFNST signaling.
\setlength{\tabcolsep}{1pt}
\begin{table}
    \renewcommand{\arraystretch}{2}
	\caption{Definition of the index \textit{trPairIdx} linking $\mathbf{U}$ to the implicit LFNST signaling. In the row \enquote{Wide angle intra mode index}, each range between brackets refers a group of indices of VVC intra prediction modes, excluding the MIP modes. $\left(*\right)$ denotes $\left\{ 0, 1 \right\} \cup$ the MIP modes indices. $\left(**\right)$ denotes $\left[ \vert -14, -1 \vert \right] \cup \left[ \vert 2, 12 \vert \right]$. The \enquote{Transform set index} indexes the pairs of LFNST matrices.}
	\centering
	\begin{tabular}{c|c|c|c|c|c|c|c}
	    \hline
	    \makecell{wide angle intra\\mode index} & $\left(*\right)$ & $\left(**\right)$ & $\left[ \vert 13, 23 \vert \right]$ & $\left[ \vert 24, 34 \vert \right]$ & $\left[ \vert 35, 44 \vert \right]$ & $\left[ \vert 45, 55 \vert \right]$ & $\left[ \vert 56, 83 \vert \right]$\\
		\hline
		\makecell{transform\\set index \cite{low_frequency_non_separable}} & $0$ & $1$ & $2$ & $3$ & $3$ & $2$ & $1$\\
		\hline
		\makecell{transposition\\of the primary\\transform coeffs} & false & false & false & false & true & true & true\\
		\hline
		\textit{trPairIdx} & $0$ & $1$ & $2$ & $3$ & $4$ & $5$ & $6$\\
		\hline
	\end{tabular}
	\label{table:indexing_implicit_lfnst}
\end{table}

As seven different pairs of transformations of the current primary transform coefficients before their potential transposition into secondary transform coefficients are actually possible, see Table \ref{table:indexing_implicit_lfnst}, the index \textit{trPairIdx} $\in \left[ \vert 0, 6 \vert \right]$ is introduced to connect $\mathbf{U}$ to the implicit LFNST signaling. Moreover, the constraint of following the same implicit signaling for the different values \textit{lfnstIdx} $\in \left\{ 1, 2 \right\}$ can be removed to gain flexibility. Then, \textit{trPairIdx} is duplicated into $\left\{ \text{\textit{trPairIdx}}_{\text{\textit{lfnstIdx}}} \right\}_{\text{\textit{lfnstIdx}} \in \left\{ 1, 2 \right\}}$.

\begin{figure}
	\centering
	\includegraphics[width=\linewidth]{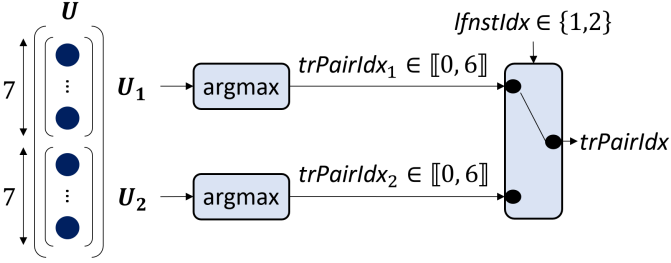}
	\caption{Computation of the index \textit{trPairIdx} of the pair of secondary transforms selected according to the neural network model from $\mathbf{U}$.}
	\label{figure:modified_indexing}
	\vspace{-4.0mm}
\end{figure}

Given the new indexing, at the VVC encoder, for a given $w \times h$ block, the neural network $f_{h, w} \left( \; . \; ; \boldsymbol{\theta}_{h, w} \right)$ computes from the pre-processed version of the context of this block the unscaled log-probability of each of the seven pairs of secondary transforms for \textit{lfnstIdx} $= 1$ and the unscaled log-probability of each of the seven pairs of secondary transforms for \textit{lfnstIdx} $= 2$. Then, for each \textit{lfnstIdx} $\in \left\{ 1, 2 \right\}$, $\text{\textit{trPairIdx}}_{\text{\textit{lfnstIdx}}}$ is the position of the largest unscaled log-probability. Finally, the selected pair index \textit{trPairIdx} in $\left\{ \text{\textit{trPairIdx}}_{1}, \text{\textit{trPairIdx}}_{2} \right\}$ depends on the value of \textit{lfnstIdx} found by the encoder. Note that, at the VVC decoder, the same procedure applies, except that the value of \textit{lfnstIdx} is read from the bitstream. Thus, for LFNST, the top-right part of Figure \ref{figure:intra_prediction_implicit_signaling} becomes Figure \ref{figure:modified_indexing}.

\subsection{Training of the neural network-based LFNST selection} \label{subsection:training_of_the_neural}
During the training of the neural network-based LFNST selection, as two learned LFNST selections for each value \textit{lfnstIdx} $\in \left\{ 1, 2 \right\}$ are considered separately, for each training example, the objective function should involve the ground truth selected secondary transform indices for \textit{lfnstIdx} $= 1$ and \textit{lfnstIdx} $= 2$ respectively as classification labels. Therefore, the objective function $\mathcal{L} \left( S_{h, w}; \boldsymbol{\phi}_{h, w} \right)$ to be minimized over the parameters $\boldsymbol{\phi}_{h, w}$ in the branch of $f_{h, w} \left( \; . \; ; \boldsymbol{\theta}_{h, w} \right)$ dedicated to the LFNST selections is expressed as
\begin{equation}
\begin{split}
	&\mathcal{L} \left( S_{h, w}; \boldsymbol{\theta}_{h, w} \right) = \frac{1}{N} \sum\limits_{\left( \mathbf{X}_{c}, i_{1}, i_{2} \right) \in S_{h, w}} - \mathcal{H} \left( \mathbf{X}_{c}, i_{1}, i_{2} ; \boldsymbol{\theta}_{h, w} \right)\\
	&\mathcal{H} \left( \mathbf{X}_{c}, i_{1}, i_{2} ; \boldsymbol{\theta}_{h, w} \right) = \log \left( \sigma \left( \mathbf{U}\left[ 0:7 \right] \right)_{i_{1}} \right) + \log \left( \sigma \left( \mathbf{U}\left[ 7: \right] \right)_{i_{2}} \right)\\
	&\left\{ \hat{\mathbf{Y}}_{c}, \mathbf{U} \right\} = f_{h, w} \left( \mathbf{X}_{c} ; \boldsymbol{\theta}_{h, w} \right) \;\; \text{and} \;\; N = \text{card} \left( S_{h, w} \right) \nonumber
\end{split}
\end{equation}
where $S_{h, w}$ denotes the training set of triplets of the pre-processed version $\mathbf{X}_{c}$ of the context of a $w \times h$ block, the index $i_{1}$ of the selected secondary transform for \textit{lfnstIdx} $= 1$ when encoding this block via VVC, and the index $i_{2}$ of the selected secondary transform for \textit{lfnstIdx} $= 2$. $\sigma$ denotes the softmax. Note that the array indexation $\mathbf{U} \left[ 0:7 \right]$ excludes the coefficient of $\mathbf{U}$ of index $7$, as in C++. The training hyperparameters will be detailed right before the experiments in Section \ref{subsection:rate_distortion_analysis_of} as they depend on the chosen neural network architectures.

\subsection{Signaling in VVC} \label{subsection:signaling_in_vvc}
Up to now, for a CB predicted via the neural network-based intra prediction mode, the index \textit{trPairIdx} of the pair of secondary transforms selected according to the neural network model has directly specified the implicit LFNST signaling. This is called the \enquote{inference} scheme.

\begin{figure}
	\centering
	\includegraphics[width=\linewidth]{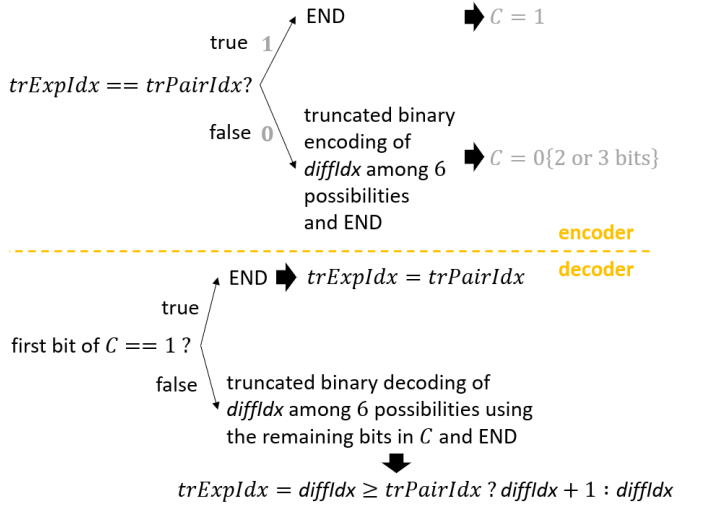}
	\caption{Predictive encoding and decoding of the pair index \textit{trExpIdx} with respect to its prediction \textit{trPairIdx} from the neural network model. $C$ is the code of the remainder of the predictive encoding written to the bitstream.}
	\label{figure:predictive_coding}
	\vspace{-4.0mm}
\end{figure}

Alternatively, a predictive coding, called \enquote{prediction} scheme, can be constructed from the learned LFNST selection. In this case, a new syntax element \textit{trExpIdx} replaces \textit{trPairIdx} in Table \ref{table:indexing_implicit_lfnst}. \textit{trPairIdx} in Figure \ref{figure:modified_indexing} becomes a prediction of \textit{trExpIdx}. At the VVC encoder, for a CB predicted via the neural network-based mode, for \textit{lfnstIdx} $\neq 0$, the best value of \textit{trExpIdx} in terms of rate-distortion is found, and the remainder of the predictive coding of \textit{trExpIdx} with respect to \textit{trPairIdx} is written to the bitstream. Figure \ref{figure:predictive_coding} details the predictive coding used in this work.

Note that, in both the \enquote{inference} and \enquote{prediction} schemes, for a CB predicted by an intra prediction mode different from the single additional neural network-based intra prediction mode, the transform signaling in VVC remains unchanged.

% Experiments.
% !TeX root = combined_neural_network.tex

\section{Experiments} \label{section:experiments}
Now that the proposed learned LFNST selection is specified, its relevance in terms of rate-distortion in VVC can be studied, see Section \ref{subsection:rate_distortion_analysis_of}. Then, our single neural network-based intra prediction mode with the learned LFNST selection inside VVC is compared to the state-of-the-art, see Section \ref{subsection:comparison_to_the}.

\subsection{Rate-distortion analysis of the learned LFNST selection} \label{subsection:rate_distortion_analysis_of}
If, as said in the second paragraph of Section \ref{section:neural_network_based_transform}, in the neural network-based mode, blocks of each possible size in VVC are predicted by a different neural network, the number of neural networks integrated into VVC would be large, and, as a common neural network for intra prediction may contain numerous parameters, the neural network parameters would incur an excessive memory footprint. To circumvent this, only the blocks of each size in $\left\{ 4 \times 4, 8 \times 4, 16 \times 4, 32 \times 4, 8 \times 8, \right.$ $\left. 16 \times 8, 16 \times 16, 32 \times 32 \right\}$ are predicted by a different neural network in the single additional mode. The single additional mode thus comprises $8$ neural networks. To maximize the usage rate of the neural network-based mode, the following steps are added. The context $\mathbf{X}$ of a $32 \times 16$ block is downsampled horizontally by $2$ before the pre-processing step in Figure \ref{figure:intra_prediction_implicit_signaling} and the neural network prediction after the post-processing step is interpolated horizontally by $2$, making the prediction of this block via $f_{16, 16} \left( \; . \; ; \boldsymbol{\theta}_{16, 16} \right)$ feasible. The same goes for a $64 \times 64$ block, but the horizontal and vertical downsampling and interpolation factors are $2$, and $f_{32, 32} \left( \; . \; ; \boldsymbol{\theta}_{32, 32} \right)$ is used for prediction. Besides, for $\left( h, w \right) \in \left\{ \left( 8, 4 \right), \left( 16, 4 \right), \left( 32, 4 \right)\right.$ $\left.\left( 16, 8 \right), \left( 32, 16 \right) \right\}$, the context $\mathbf{X}$ of a $w \times h$ block is transposed before the pre-processing step and the neural network prediction after the post-processing step is transposed, allowing the prediction of this block via $f_{w, h} \left( \; . \; ; \boldsymbol{\theta}_{w, h} \right)$.

From now on, the following parametrization of the context of a $w \times h$ block applies. If $\min \left( h, w \right) \leq 8$, $n_{a} = n_{l} = \min \left( h, w \right)$. Otherwise, $n_{a} = h \mathbin{/} 2$ and $n_{l} = w \mathbin{/} 2$. Moreover, the pre-processing and post-processing steps in Figure \ref{figure:intra_prediction_implicit_signaling} correspond to those detailed in \cite{iterative_training_of_neural}.

For the training in Section \ref{subsection:training_of_the_neural}, the RGB images in the ILSVRC 2012 training dataset and those in DIV2K converted into $\text{Y}\text{C}_{\text{b}}\text{C}_{\text{r}}$ are encoded via VTM-8.0 with Quantization Parameter (QP) drawn from $\left\{ 22, 27, 32, 37 \right\}$ for each image. Each neural network training runs for $800000$ iterations with batch size $100$, ADAM, and $0.0002$ as learning rate.

As this work does not relate to the enhancement of the neural network prediction of a block, the objective function on block prediction is picked from \cite{iterative_training_of_neural} and the neural network architectures in \cite{iterative_training_of_neural} are simply adapted to the learned LFNST selection. Moreover, the architectures are reduced to decrease the modified VVC encoder and decoder running times, see Tables \ref{table:architecture_fully_connected} to \ref{table:architecture_merger}. Note that, when $\min \left( h, w \right) > 8$, $\mathbf{X}_{c}$ is split into two portions, see \cite{iterative_training_of_neural}, each portion being fed into a different convolutional branch of $f_{h, w} \left( \; . \; ; \boldsymbol{\theta}_{h, w} \right)$.
\begin{table}
	\caption{Architecture of $f_{h, w} \left( \; . \; ; \boldsymbol{\theta}_{h, w} \right)$ where $\min \left( h, w \right) \leq 8$. In the column \enquote{input}, a number refers to the index of the layer whose output is the input to the current layer. The layers of indices $3$ and $4$ return $\hat{\mathbf{Y}}_{c}$ and $\mathbf{U}$ respectively.}
	\centering
	\begin{tabular}{c|c|c|c|c}
	    \hline
	    layer index & input & layer type & number of neurons & non-linearity\\
	    \hline
	    $1$ & $\mathbf{X}_{c}$ & fully-connected & $1200$ & LeakyReLU\\
		\hline
		$2$ & $1$ & fully-connected & $1200$ & LeakyReLU\\
		\hline
		$3$ & $2$ & fully-connected & $hw$ & -\\
		\hline
		$4$ & $2$ & fully-connected & $14$ & -\\
		\hline
	\end{tabular}
	\label{table:architecture_fully_connected}
	\vspace{-1.0mm}
\end{table}
\begin{table}
	\caption{Architecture of the branch of $f_{16, 16} \left( \; . \; ; \boldsymbol{\theta}_{16, 16} \right)$ taking the above portion $\mathbf{X}_{0}$ of $\mathbf{X}_{c}$. For $f_{32, 32} \left( \; . \; ; \boldsymbol{\theta}_{32, 32} \right)$, the same architecture applies but the strides in bold become $\left( 2, 2 \right)$.}
	\centering
	\begin{tabular}{c|c|c|c|c|c|c}
	    \hline
	    layer index & input & layer type & filter size & nb of filters & stride & non-linearity\\
	    \hline
	    $1$ & $\mathbf{X}_{0}$ & convolutional & $3 \times 3 \times 1$ & $32$ & $\left( 2, 2 \right)$ & LeakyReLU\\
		\hline
		$2$ & $1$ & convolutional & $3 \times 3 \times 32$ & $64$ & $\left( 2, 2 \right)$ & LeakyReLU\\
		\hline
		$3$ & $2$ & convolutional & $3 \times 3 \times 64$ & $128$ & $\left( \mathbf{1}, \mathbf{2} \right)$ & LeakyReLU\\
		\hline
		$4$ & $3$ & convolutional & $3 \times 3 \times 128$ & $128$ & $\left( 1, 2 \right)$ & LeakyReLU\\
		\hline
		$5$ & $4$ & flattening & - & - & - & -\\
		\hline
	\end{tabular}
	\label{table:architecture_branch_above}
	\vspace{-1.0mm}
\end{table}
\begin{table}
	\caption{Architecture of the branch of $f_{16, 16} \left( \; . \; ; \boldsymbol{\theta}_{16, 16} \right)$ taking the left portion $\mathbf{X}_{1}$ of $\mathbf{X}_{c}$. For $f_{32, 32} \left( \; . \; ; \boldsymbol{\theta}_{32, 32} \right)$, the same architecture applies but the strides in bold become $\left( 2, 2 \right)$.}
	\centering
	\begin{tabular}{c|c|c|c|c|c|c}
	    \hline
	    layer index & input & layer type & filter size & nb of filters & stride & non-linearity\\
	    \hline
	    $6$ & $\mathbf{X}_{1}$ & convolutional & $3 \times 3 \times 1$ & $32$ & $\left( 2, 2 \right)$ & LeakyReLU\\
		\hline
		$7$ & $6$ & convolutional & $3 \times 3 \times 32$ & $64$ & $\left( 2, 2 \right)$ & LeakyReLU\\
		\hline
		$8$ & $7$ & convolutional & $3 \times 3 \times 64$ & $128$ & $\left( \mathbf{2}, \mathbf{1} \right)$ & LeakyReLU\\
		\hline
		$9$ & $8$ & convolutional & $3 \times 3 \times 128$ & $128$ & $\left( 2, 1 \right)$ & LeakyReLU\\
		\hline
		$10$ & $9$ & flattening & - & - & - & -\\
		\hline
	\end{tabular}
	\label{table:architecture_branch_left}
	\vspace{-2.0mm}
\end{table}
\begin{table}
	\caption{Architecture of the part of $f_{16, 16} \left( \; . \; ; \boldsymbol{\theta}_{16, 16} \right)$ merging the outputs of the two branches in Tables \ref{table:architecture_branch_above} and \ref{table:architecture_branch_left}. The layers of indices $13$ and $14$ return $\hat{\mathbf{Y}}_{c}$ and $\mathbf{U}$ respectively. For $f_{32, 32} \left( \; . \; ; \boldsymbol{\theta}_{32, 32} \right)$, the same architecture applies but the layer of index $12$ contains $hw$ neurons instead of $500$.}
	\centering
	\begin{tabular}{c|c|c|c|c}
	    \hline
	    layer index & input & layer type & number of neurons & non-linearity\\
	    \hline
	    $11$ & $5$ and $10$ & concatenation & - & -\\
	    \hline
	    $12$ & $11$ & fully-connected & $500$ & LeakyReLU\\
		\hline
		$13$ & $12$ & fully-connected & $hw$ & -\\
		\hline
		$14$ & $12$ & fully-connected & $14$ & -\\
		\hline
	\end{tabular}
	\label{table:architecture_merger}
	\vspace{-2.0mm}
\end{table}

Finally, for a given CB, the intra signaling of the neural network-based mode explained in \cite{iterative_training_of_neural} is re-used here.
\setlength{\tabcolsep}{8pt}
\begin{table*}
	\caption{Mean BD-rate reductions in $\%$ of VTM-8.0 with the single additional neural network-based mode w.r.t VTM-8.0. Only the first frame of each sequence in the JVET CTC \cite{jvet_common_test} is considered. The largest absolute mean BD-rate reduction in luminance is in bold.}
	\centering
	\begin{tabular}{lcccccccccccc}
		\hline
		\multirow{2}{*}{Video class} & \multicolumn{3}{c}{default} & \multicolumn{3}{c}{fully explicit LFNST} & \multicolumn{3}{c}{inference} & \multicolumn{3}{c}{prediction}\\ \cline{2-13} & Y & $\text{C}_{\text{b}}$ & $\text{C}_{\text{r}}$ & Y & $\text{C}_{\text{b}}$ & $\text{C}_{\text{r}}$ & Y & $\text{C}_{\text{b}}$ & $\text{C}_{\text{r}}$ & Y & $\text{C}_{\text{b}}$ & $\text{C}_{\text{r}}$\\
		\hline
		A1 & $-4.41$ & $-3.78$ & $-3.45$ & $-5.04$ & $-4.36$ & $-4.35$ & $-5.23$ & $-4.37$ & $-4.59$ & $\mathbf{-5.46}$ & $-4.66$ & $-4.74$\\
		\hline
		A2 & $-2.02$ & $-1.58$ & $-1.91$ & $-2.35$ & $-2.00$ & $-2.29$ & $-2.51$ & $-2.48$ & $-2.13$ & $\mathbf{-2.55}$ & $-2.08$ & $-2.62$\\
		\hline
		B & $-2.54$ & $-2.35$ & $-2.07$ & $-3.03$ & $-2.27$ & $-2.20$ & $-3.15$ & $-2.40$ & $-2.83$ & $\mathbf{-3.31}$ & $-2.78$ & $-2.98$\\
		\hline
		C & $-2.53$ & $-1.74$ & $-2.01$ & $-2.88$ & $-2.76$ & $-2.79$ & $-3.04$ & $-2.05$ & $-2.16$ & $\mathbf{-3.06}$ & $-2.75$ & $-2.65$\\
		\hline
		D & $-2.66$ & $-2.38$ & $-1.22$ & $-3.19$ & $-2.78$ & $-2.72$ & $-3.25$ & $-3.05$ & $-2.57$ & $\mathbf{-3.42}$ & $-3.46$ & $-2.51$\\
		\hline
		E & $-3.99$ & $-3.32$ & $-3.79$ & $-4.48$ & $-3.23$ & $-3.81$ & $-4.46$ & $-3.87$ & $-4.71$ & $\mathbf{-4.66}$ & $-3.97$ & $-4.38$\\
		\hline
		F & $-1.46$ & $-1.58$ & $-1.90$ & $-1.68$ & $-1.35$ & $-1.50$ & $-1.71$ & $-2.03$ & $-2.01$ & $\mathbf{-1.97}$ & $-2.02$ & $-1.73$\\
		\hline
		Mean & $-3.01$ & $-2.49$ & $-2.55$ & $-3.46$ & $-2.84$ & $-2.97$ & $-3.58$ & $-2.91$ & $-3.17$ & $\mathbf{-3.71}$ & $-3.17$ & $-3.37$\\
		\hline
	\end{tabular}
	\label{table:analysis_learned_lfnst}
	\vspace{-4.0mm}
\end{table*}

To assess the relevance of the proposed learned LFNST selection, the \enquote{inference} and \enquote{prediction} schemes must be compared in terms of rate-distortion against a baseline in which, for a given CB predicted via the neural network-based mode, if \textit{lfnstIdx} $\in \left\{ 1, 2 \right\}$, the pair of LFNST matrices of transform set index $0$ is always chosen, without transposing the primary transform coefficients. This is called the \enquote{default} scheme. Another interesting baseline, called \enquote{fully explicit LFNST} corresponds to the \enquote{prediction} scheme without the neural network-based prediction of \textit{trExpIdx}. This means that, at the VVC encoder, for a given CB predicted by the neural network-based mode, for \textit{lfnstIdx} $\in \left\{ 1, 2 \right\}$, the best value of \textit{trExpIdx} in terms of rate-distortion is found, and this value is written to the bitstream via a truncated binary encoding. In the experiments, only the first frame of each video sequence of the JVET CTC \cite{jvet_common_test} is considered. The configuration is all-intra.

The most striking remark is that the \enquote{default} scheme is much worse than the three other transform selection schemes in terms of rate-distortion, see Table \ref{table:analysis_learned_lfnst}. Besides, on the luminance channel, the \enquote{prediction} scheme adds $-0.25\%$ of mean BD-rate reduction with respect to the \enquote{fully explicit LFNST}. The \enquote{inference} scheme yields $-0.12\%$ of additional mean BD-rate reduction with respect to the \enquote{fully explicit LFNST}. These three observations prove the relevance of the proposed neural network-based transform selections. As the \enquote{inference} scheme adds no explicit signaling to VVC, on the luminance channel, the difference of $-0.57\%$ in mean BD-rate reduction between the \enquote{inference} scheme and the \enquote{default} scheme can be viewed as the pure rate-distortion gain of the neural network-based transform selections. This gain arises from bit-rate drop at close PSNRs. Table \ref{table:analysis_complexity} reports the mean encoder and decoder running times of each tested scheme.

Note that an approach involving different neural networks for each QP (or range of QPs) was not studied in this work as this type of approach increases significantly the memory footprint of the neural network parameters inside VVC.
\setlength{\tabcolsep}{6pt}
\begin{table}
	\caption{Mean encoder and decoder running times of VTM-8.0 including the single additional neural network-based mode with respect to VTM-8.0 on the JVET CTC in all-intra. $100\%$ means the same running time as VTM-8.0.}
	\centering
	\begin{tabular}{lcccc}
		\hline
		 & default & fully explicit LFNST & inference & predictive\\
		\hline
		Encoder & $369\%$ & $387\%$ & $400\%$ & $499\%$\\
		\hline
		Decoder & $3330\%$ & $3551\%$ & $3854\%$ & $5052\%$\\
		\hline
	\end{tabular}
	\label{table:analysis_complexity}
	\vspace{-2.0mm}
\end{table}

\subsection{Comparison to the state-of-the-art} \label{subsection:comparison_to_the}
Most of the previous approaches on the neural network-based intra prediction for block-based video coding integrate their tool into HEVC \cite{progressive_spatial_recurrent_neural, fully_connected_network_based, fully_neural_network_mode, multi_scale_convolutional_neural, generative_adversarial_network_based, a_hybrid_neural}, the ancestor of VVC. For comparison, our proposed neural network-based transform selection should also be integrated into HEVC. However, this makes little sense in HEVC as, unlike in VVC, the implicit transform signaling in HEVC is extremely limited. Indeed, the DST7-DST7 applying to $4 \times 4$ luminance blocks predicted in intra instead of the DCT2-DCT2 is the main implicit transform signaling in HEVC. For a rate-distortion comparison between the neural network-based intra prediction used in this paper without the neural network-based transform selection and several neural network-based intra prediction tools in the literature, all inside HEVC, please see \cite{iterative_training_of_neural}.

To our knowledge, as of now, only \cite{chroma_intra_prediction_with} presents a neural network-based intra prediction tool tested in a recent version of VVC. A neural network-based intra prediction mode for chrominance featuring an attention mechanism is put into VTM-7.0. Using the same test data as in Table \ref{table:analysis_learned_lfnst}, $-0.15\%$, $-0.68\%$, $-0.53\%$ of mean BD-rate reduction is reported in all-intra. The mean encoder and decoder running times of their modified VTM-7.0 w.r.t VTM-7.0 are $120\%$ and $947\%$.

% Conclusion.
% !TeX root = combined_neural_network.tex

\section{Conclusion} \label{section:conclusion}
This paper has introduced a combined neural network-based intra prediction and transform selection for a block-based video codec. As the neural network-based transform selection depends on both the decoded samples around the current block and the neural network intra prediction function, it can modelize all the intra-transform correlations. When integrated into VTM-8.0, this approach yields large rate-distortion gains.

\bibliographystyle{IEEEtrans}
\bibliography{combined_neural_network}

\end{document}